\documentstyle[12pt]{article}

\begin{document}

\title{ QCD sum rules as a tool for investigation of
characteristics of a nucleon in nuclear matter}
\author{E.G. Drukarev  \\
Petersburg Nuclear Physics Institute\\
Gatchina, St.Petersburg 188350, Russia }
\date{} \maketitle

Progress of traditional nuclear physics in describing changes of
properties of nucleons in nuclear matter is acknowledged. However, this
approach is based on conception of $NN$ interactions and needs some
phenomenological input on the latter. The small distances of the order
of the nucleon radius appear to be very important. Also, calculation of
each new characteristics requires further development of theory. Say,
in order to obtain neutron-\-proton mass difference one should develop
theory of charge-\-symmetry breaking forces. Lastly, there are the
problems which are inaccessible for traditional nuclear physics.
Swelling of nucleon in nuclei, demonstrated by EMC group is an example.

This stimulates investigation of alternative approach. The application
of QCD sum rules (SR) at finite density was suggested in [1]. The QCD
SR in vacuum [2] succeeded earlier in describing properties of free
hadrons. The method is based on dispersion relations for the function
(correlator) describing propagation of the system carrying the quantum
numbers of the nucleon. Expansion in inverse powers of square momentum
$q^2$ is related to observable characteristics of the nucleon. Hence,
the latter are expressed through the values of QCD condensates. The
method is based on QCD Lagrangian, employes crucially the properties of
strong interactions at small distances and contains confinement as an
input.

The SR in nuclear matter tie the changes of the values of certain QCD
condensates with characteristics of nucleons in the medium.
Single-\-particle potential energy appears to be the sum of the terms
proportional to expectation values of quark operators $\bar q\gamma_0q$
and $\bar qq$ [1]. The former one vanishes in vacuum while the vacuum
value should be subtracted in the latter case. The vector term is
exactly linear in density, providing contribution of the order+250 MeV.
The scalar one is about ($-300$ MeV). Hence, the method reproduces the
main points of relativistic nuclear physics. It provides also
connection between the scalar fields and pion-\-nucleon sigma term.

The neutron-proton mass difference was found to contain dependence on
isospin breaking expectation value of the operator $\bar dd-\bar uu$
[3]. Neutron was bound to be found stronger than the proton with
reasonable value of the mass difference. The charge symmetry breaking
in the scalar channel was shown to be as important as in the vector
one. Thus the approach provides guide-\-lines for traditional nuclear
physics.

The method was applied to investigation of the deep inelastic structure
functions of nuclei [4]. The deviations from that of a system of free
nucleons were shown to be determined by the four quark condensate. The
calculated values followed typical EMC behaviour. However in the first
step, made in [4], we did not touch cumulative aspects of the problem.

The method can be used for calculation of other properties of nucleons
in medium. It can be improved within it's own framework by taking into
account higher terms of $q^{-2}$ expansion, containing  more
complicated in-\-medium condensates and by more detailed description of
spectral density.
This activity is supported by the Russian Fund for Fundamental
Research (grant\#95-02-03752-a).

\noindent 1. E.G.Drukarev, E.M.Levin, Progr.in Part \& Nucl.Phys.
{\bf22} (1991) 77.\\
2. M.A.Shifman, A.I.Vainshtein, V.I.Zakharov,
Nucl.Phys. {\bf B147} (1979) 385.\\
3. E.G.Drukarev, M.G.Ryskin,
Nucl.Phys. {\bf A572} (1994) 560; {\bf A577} (1994) 375.\\
4. E.G.Drukarev, M.G.Ryskin, Z.f.Phys.A {\bf356} (1997) 457.
\end{document}